# ENTROPIC IDENTIFICATION OF THE FIRST ORDER FREEZING TRANSITION OF A SUSPENSION OF HARD SPHERE PARTICLES

W. van Megen[1] and H. J. Schöpe[2]

[1]*Department of Applied Physics, Royal Melbourne Institute of Technology, Melbourne, Victoria 3000, Australia*
[2]*Institut für Angewandte Physik, Universität Tübingen, Auf der Morgenstelle 10, 72076 Tübingen, Germany*

ABSTRACT

We analyse the experimental particle current auto-correlation function (CAF) of suspensions of hard spheres. Interactions between the particles are mediated by thermally activated acoustic excitations in the solvent. Those acoustic modes are tantamount to the system's (energy) microstates and by their orthogonality, each of those modes can be identified with an independent Brownian particle current. Accordingly, partitioning of the system's energy states is impressed on the CAF. This impression provides a novel measure of the entropy and location of a partitioning/entropy limit at a packing fraction that coincides with that of the observed suspension's first order freezing transition.

*Introduction*. – The system of particles with hard sphere interactions has proved to be a valuable reference that has enhanced considerably our understanding of condensed matter. Its first order freezing/melting transition, in particular, has been studied extensively. It was first discovered by computer simulation in 1957 [1,2]. In 1964 its location was quantified, again by computer simulations that found the packing fractions, $\phi$, of the coexisting phases of the hard sphere fluid and crystal to be $\phi_f = 0.494$ and $\phi_m = 0.545$ [3]. Just over 20 years later experiment showed that these phase equilibria are replicated by suspensions of (colloidal) particles stabilised against coagulation by thin steric surface coatings[4]. In the absence of direct, finite range interactions among the particles the transition, expressed in thermodynamic terms, occurs when the entropy associated with the particles' free volume in the crystal exceeds the configurational entropy of the fluid [5-7].

Statistical thermodynamics is underpinned by partitioning of energy; a concept which naturally also applies to the phase equilibria. Ergo, the description of the transition must ultimately be underpinned by the manner of energy partitioning in the respective phases, in which terms the above statement reads: the fluid to crystal transition occurs when the number of distinct accessible energy states, or microstates, of the crystal exceeds that of the fluid. In formation of the crystal phase the increase in entropy is self-evident; here the (excess) entropy resides in the countable lattice modes. Moreover, as is clear from numerous studies of crystallisation kinetics of suspensions of hard spheres, for instance, development of lattice modes by whatever combination of nucleation and growth is accompanied by an increase in the particles' free volume [8-10]; here the connection between entropy and free volume is evident. While compression of the hard-sphere fluid necessarily leads to loss of configurational entropy, it is less transparent how that loss is related to, or results from, a corresponding reduction in accessible energy states. The aim of this Letter is to quantify that connection and explore its consequences.

We recognise that the freezing transition is basically structural in nature and independent of whether the intrinsic particle dynamics are ballistic or diffusive. However, in the case of a suspension of sufficiently large particles, fluctuations in their spatial configurations are vastly slower than energy exchanges among those particles. As a result, the signatures of energy partitioning turn out to be more transparent than in molecular fluids. The first simplification admitted by the separation of time scales, expressed by the Fokker-Planck equation [11-14], treats the suspending liquid (solvent) as a fluctuating hydrodynamic continuum – a momentum field, comprising propagating, longitudinal (sound) and diffusing, transverse (viscous flow) components. We consider just the acoustic modes for only these effect energy exchanges between the particles and, in the statistical thermodynamics description of this system, only these determine the partition function of the system. The next simplification considers these exchanges instantaneous on the time scales ($\tau \gtrsim 10^{-6}$ s) on which particle motions are observed in conventional optical experiments, spectroscopic or microscopic [11,15]. On the sonic time scale ($\tau_{sonic} \sim 10^{-10}$ s) – the time for sound to propagate typical distances between the particles – the particles are effectively stationary and the "instantaneous", ensemble average of their spatial distribution, expressed by the structure factor, $S(q)$, furnishes the fixed, reflecting boundary conditions that determine the frequencies of the sound modes; $S(q)$ effects partitioning of the acoustic modes in the (interstitial) solvent. In other words, the "snapshot" of the ensemble of particle positions effectively constitute a resonating acoustic box whose boundaries, however convoluted, determine (I) the number, N, of normal orthogonal – distinct and independent acoustic modes and, (II) the spatial distribution of those modes by their being pinned to the ensemble-averaged distribution of the particles and extent of their exclusion from the space occupied by the particles. Note, in particular, that (I), just the *number* of acoustic modes which, in this case, is tantamount to the number of microstates, is independent of the spatial frequency, q.

On experimental times these rapidly fluctuating, but deterministic, acoustic excitations are "visible" as Brownian motion and manifest in the time correlation functions of various randomly fluctuating quantities. The present study rests on the longitudinal current auto-correlation function (CAF), which property provides a more direct link to the energy exchanges between the particles than the more usually studied correlation function of the particle number density [11]. This is evident from previous observations that the CAF admits to a time scaling [16,17] that leaves the result independent of q. This presents the first indication of a connection between the scaled CAF and contribution (I) above: A connection with the number, N, of microstates of the system – or, in other words, the entropy, $S \sim \ln(N)$. Analysis below aims to render this connection more rigorous by identifying in the CAF conservation of energy and denumerability of the microstates.

In singling out just the longitudinal momentum current in our approach, we have stopped short of presuming that *all* momenta have relaxed to equilibrium. While this level of coarse graining, expressed by the Smoluchowski equation, is more commonly adopted in studies of the dynamics of suspensions [11-14,16], its drawback is that it precludes distinguishing the separate roles of the longitudinal and transverse momentum currents and, thereby, negating any possibility of identifying energy partitioning in the time correlation functions, which we aim to achieve here.

*Methods.* – The results comprise CAFs derived from dynamic light scattering (DLS) measurements on experimentally established hard sphere like model systems; suspensions of polymer particles, labelled P [18], and microgel particles, M1 and M2 [17]. The samples' packing fractions, $\phi$, are determined by referencing the observed equilibrium colloidal fluid-crystal phase separation, in each case, to that of the ideal hard sphere system[4,19,20]. Accordingly, the freezing value is set at $\phi_f$=0.494. Properties of the particles immediately relevant for the present study are summarized in Table 1. Other properties of the suspensions and light scattering procedures are documented elsewhere [10,18,19]. In the results presented below the spatial frequency, q, and all lengths are expressed in terms of the particle radius, R (Table 1).

|  | Suspension P | Suspension M1 | Suspension M2 |
|---|---|---|---|
| Radius, R | 185 nm | 430 nm | 370 nm |
| Polydispersity | 8 % | 4 % | 2 % |
| Brownian time, $\tau_B$ | 0.013 s. | 0.175 s. | 0.111 s |

Table 1. Suspension properties; Particle radii, polydispersities, $\sqrt{\langle R^2\rangle/\langle R\rangle^2 - 1}$ and $\tau_B = R^2/(6D_0)$, the time that characterises Brownian motion and $D_0$ the diffusion coefficient of freely diffusing particles.

As mentioned in the *Introduction* we consider the time-scaled time correlation function,

$$C^*(q,\tau^*) = -d^2 f(q,\tau^*)/d\tau^{*2} = q^2 \langle j(q,0) j^\dagger(q,\tau^*)\rangle/\langle|\rho(q)|^2\rangle, \tag{1}$$

of the longitudinal particle current density [21],

$$j(q,t) = N^{-1} \sum_{k=1}^{N} \hat{\mathbf{q}} \cdot \mathbf{v}_k(t) \exp[-i\mathbf{q} \cdot \mathbf{r}_k(t)]. \tag{2}$$

This has been derived by numerically differentiating the measured correlation function,

$$f(q,\tau^*) = \langle\rho(q,0)\rho^\dagger(q,\tau^*)\rangle/\langle|\rho(q)|^2\rangle, \tag{3}$$

of the particle concentration, $\rho(t)$. "$\dagger$" indicates the complex conjugate and $\hat{\mathbf{q}}=\mathbf{q}/q$ is the unit propagation vector. The delay time, $\tau$, usually expressed in terms of the Brownian time, $\tau_B$ (Table 1), is further scaled here by the time, $1/q^2 D(q)$, that characterises the diffusive decay of concentration fluctuations of spatial frequency q: ie,

$$\tau^* = q^2 D(q)\tau. \tag{4}$$

Here $D(q) = D_0 H(q)/S(q)$ denotes the short time diffusion coefficient and $H(q)$ the hydrodynamic factor [11].

*Results and analyses.* – The results of experiments on the three suspensions (Table 1) are quantitatively consistent regardless of the differences in their chemical composition, particle radii and polydispersities. Since the time-scaled CAFs, $C^*(q,\tau^*)$, of P and M1 have been published previously [16,17], it suffices to show just a representative result, at $\phi$=0.351, for suspension M2 in Fig. 1a. This illustrates the key feature, found for *all* packing fractions of the three suspensions in the equilibrium fluid phase ($\phi<\phi_f$), that $C^*(q,\tau^*)$ shows no systematic variation with q in the experimental windows of spatial frequency, $1 \lesssim q \lesssim 5$, that bracket the position, $q_m$, of the primary maximum in S(q), and delay time, $-1.5 \lesssim \log\tau^* \lesssim 1.5$ [outside this space-time window experimental noise precludes any inference]. Absolute values are plotted because $C(q,\tau^*)$, being constrained by conservation of particle number density, decays from below.

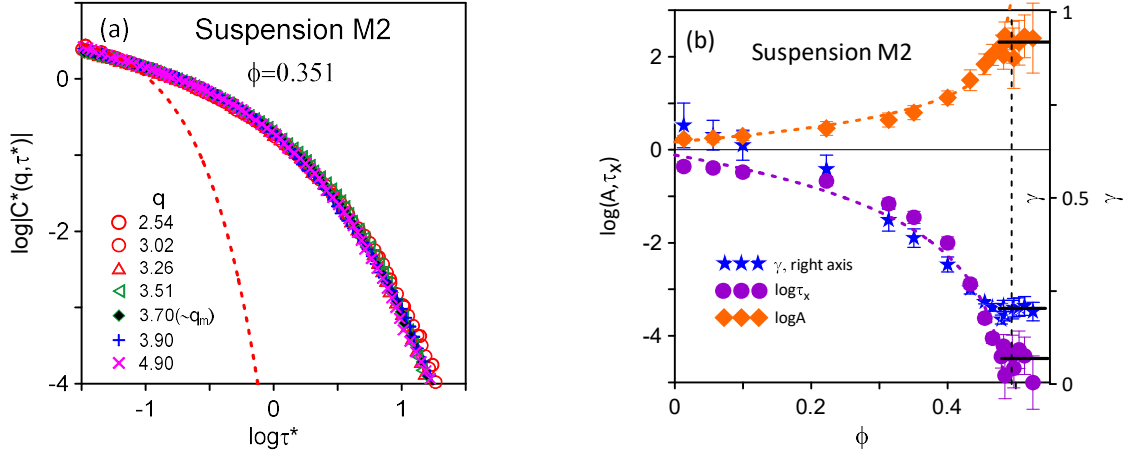

Fig. 1. (a) The scaled current correlator for suspension M2 for values of q indicated. The best fitting SE, Eq. (5), (A=0.7, $\tau_x$=0.07, $\gamma$=0.44) cannot be distinguished from the data. To illustrate the degree of stretching the basic exponential decay (SE with $\gamma$=1) is shown by the red dashed curve. (b) Parameters A, $\tau_x$ and $\gamma$ of the SE defined by Eq. (5). The vertical dashed line is located at $\phi_f$=0.494. The two dashed curves are power laws $A=A_o|\phi_a - \phi|^{-\alpha}$ (top) and $\tau_x = T_o|\phi_b - \phi|^\beta$ fitted to A and $\tau_x$, respectively. Fitting parameters are listed in Table 2. The small (black) horizontal bars at the right of the figure indicate average values of the SE parameters fitted to $C^*(q\neq q^*,\tau^*;\phi>\phi_f)$. See text for explanation.

We proceed by repeating, in part, the analysis described in Ref. [16,17]; First, we approximate the scaled CAFs, by a stretched exponential (SE) function,

$$C^*(q,\tau^*) = A \exp[-(\tau^*/\tau_x)^\gamma], \qquad (5)$$

of the (scaled) delay time (Fig. 1). The fitting parameters in Fig. 1b show that the amplitude, A, increases and the characteristic decay time, $\tau_x$, and stretching index, $\gamma$, decrease with $\phi$.

Second, A and $\tau_x$ are fitted to power laws

$$A_o |\phi_a-\phi|^{-\alpha} \text{ and } T_o |\phi_b-\phi|^\beta, \qquad (6)$$

respectively (Fig. 1b). From the fit parameters, given in Table 2, one sees that in all cases $\phi_a$ and $\phi_b$ are consistent, within experimental error, with the freezing value, $\phi_f$=0.494.

In the third step in the analysis, not previously advanced, the SE is expressed by a superposition of exponential decays

$$\exp[-(\tau^*/\tau_x)^\gamma] = \int g(\tau_t)\exp[-\tau^*/\tau_t]d\tau_t. \qquad (7)$$

The moments of the distribution, $g(\tau_t)$, of decay times are defined by $<\tau_t^n> = \tau_x^n \Gamma(n/\gamma)/(\gamma\Gamma(n))$ ($\Gamma(y)$ is the Gamma function) [22]. Whether the data supports this decomposition is one of the main issues discussed below. In any case, from the parameters, $\tau_x$ and $\gamma$, we determine the average decay time, $<\tau_t>$, the normalised spread, $\sigma = (<\tau_t^2> - <\tau_t>^2)^{1/2}/<\tau_t>$, and skewness, $\zeta = (<\tau_t^3> - 3<\tau_t^2><\tau_t> + 2<\tau_t>^3)/\sigma^3$, shown in Fig. 2a for all three suspensions. As is evident from the spread in the results, these derived quantities suffer accumulation of the errors in $\tau_x$ and $\gamma$. So, power laws were not fitted the moments as was done for A and $\tau_x$ in Fig. 1b. Nonetheless, it is noteworthy that both the spread and, in particular, the *positive* skewness increase with $\phi$.

Products, $A\tau_x$ and $A<\tau_t>$, are plotted in Fig. 2b. Again, errors notwithstanding, it is evident that $A\tau_x$ decreases appreciably with $\phi$ while $A<\tau_t>$ does not.

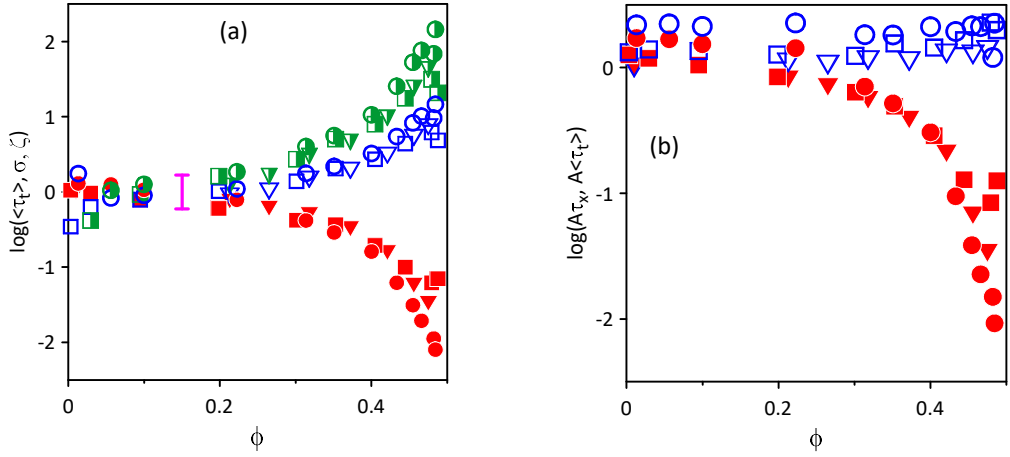

Fig. 2. Triangles: suspension P; squares, suspension M1; circles, suspension, M2. (a) Moments of the distribution, $g(\tau_t)$, of decay times in Eq. (7); Mean, $<\tau_t>$ (closed symbols). Spread, $\sigma$ (open symbols). Skewness, $\zeta$ (half-filled symbols). The (magenta) vertical bar is an estimate of the error in $<\tau_t>$. (b) Products, $A\tau_x$, (closed symbols) and $A<\tau_t>$, (open symbols). See text for details.

| Suspension | $A_o$ | $\phi_a$ | $\alpha$ | $T_o$ | $\phi_b$ | $\beta$ |
|---|---|---|---|---|---|---|
| P | 0.3 | 0.492±0.004 | 1.2 | 10 | 0.490±0.004 | 2.3 |
| M1 | 0.4 | 0.497±0.008 | 1.2 | 7 | 0.498±0.007 | 1.9 |
| M2 | 0.6 | 0.496±0.004 | 1.4 | 6 | 0.496±0.004 | 3.0 |

Table 2. Values of parameters of the power laws (Eq. 6) shown in Fig. 1b. Fits were done by the method of Gaussian weighted least squares. Errors in Table 2 reflect the uncertainties of the fit parameters.

Although our focus is primarily on the thermodynamically stable suspension ($\phi < \phi_f$), we point out that the CAFs of the metastable suspension ($\phi > \phi_f$) differ. For this case previous experiments [16,17,23] show deviations from the above time scaling and the SE approximation for spatial frequencies q=q* around $q_m$. However, for q≠q* scaling still holds and, as far as experimental noise allows us to determine, the parameters of the SE fitted to $C^*(q≠q^*,τ^*;ϕ>ϕ_f)$ are the same as those of the SE fitted to $C^*(q,τ^*;ϕ_f)$. Values of these parameters for $ϕ>ϕ_f$ for suspension M2 are included in Fig. 1b and for suspensions P and M1 presented in Ref. [23] and [17].

*Discussion.* – The implication of the observation (Fig. 1a) that, for $ϕ<ϕ_f$, the time-scaled current correlator, $C^*(q,τ^*)$, carries no explicit dependence on q, is that all interactions among the particles are captured by the short-time diffusion coefficient, D(q). For the case of suspended hard spheres this means that hydrodynamic interactions are established on the experimental time scale ($≳10^{-6}$s). Put another way, the scaling of $C^*(q,τ^*)$ is an expression or consequence of the point made in the *Introduction*, and verified in other work [24-26], that energy exchanges among the particles occur on sonic times ($τ_{sonic}$~$10^{-10}$s.). Thus, for longer delay times, having lost memory of those exchanges, the observed particle current, **j**(q,t), comprises a superposition of randomly fluctuating, complex Gaussian variables (Eq. (2)). Alternatively, we reason that each of the N mutually orthogonal acoustic modes, $ε_t$, that effects energy exchanges between the particles at sonic speed creates, in the ensemble of thermally induced initial conditions, by virtue of the Central Limit Theorem, a random Gaussian current whose time correlation function decays exponentially with decay time, $τ_t$. The time correlation function of the current (Eq. (1)) is the corresponding superposition of these exponential decays (Eq. (7)). Prior to application of the time scaling by Eq. (4), decay of the CAF can only be effected by the processes defined in the *Introduction*: (I) Random sampling of N microstates – the statistical expression, as observed on experimental times, of equilibration of the suspension that actually occurs on sonic times. This process is modulated by (II), rearrangement of particle configurations by diffusive pathways whose spatial distribution, D(q), is set by the structure S(q). Evidently, by results in Fig. 1 and those in Ref. [16,17], the time scaling effectively divides out (II), the spatial modulation imposed by the structure, and leaves in $C^*(q,τ^*)$ a decay independent of q. Whether $C^*(q,τ^*)$ in fact expresses process (I) remains to be verified. Before proceeding we emphasise that energy conservation is implicit in the observed dynamics. Thus, appeal to further interactions occurring on experimental times, in order to explain some aspect of the observed dynamics, would be inconsistent with energy conservation.

On the basis of random sampling of equally accessible microstates, a decrease in N, consequent on an increase in ϕ, results in a corresponding decrease in relaxation time $τ_x$. The decrease in $τ_x$, seen in Fig. 1b, appears to be consistent with this. In addition, conservation of (average) energy, E, stored in the acoustic modes demands that any decrease in their number, N, be compensated by a proportional increase in their average squared amplitude, $a^2$. Since the latter is proportional to the amplitude, A, of the scaled CAF, E=N*$a^2$=N*A is constant. But, as already noted, A$τ_x$ (Fig. 2b) decreases systematically with ϕ while A<$τ_t$>, on the other hand, shows no such variation. The difference is significant for it shows the product, A<$τ_t$>, is consistent with energy conservation while the other product, A$τ_x$, is not. The implication is that <$τ_t$>, rather than $τ_x$, is a measure of the average time required to randomly sample the system's N microstates; accordingly <$τ_t$>~N and log(<$τ_t$>) is a measure of the entropy. This result corroborates the inference above, the same as that in Ref. [27], albeit in another context, that absorption of all q-dependence by the time scaling (Eq. (4)) is sufficient for the superposition (Eq. (7)) to apply; the CAF can be expressed by the sum of *independent* exponential relaxation functions and its stretching attributed to a distribution, $g(\tau_t)$, in the decay times. Alternatively, as illustrated in Fig. 2b, describing the CAF by a single decay time, $\tau_x$, and attributing stretching to collective dynamics violates energy conservation.

Therefore, we now identify the spread, $\sigma$, and specifically the positive skewness, $\zeta$ (Fig. 2a), indicative of a "tail" of long decay times in the distribution, $g(\tau_t)$, of those decay times, with the tendency of the distribution of the amplitudes of the thermally excited acoustic modes to be biased toward lower frequencies. In other words, the spectrum of the momentum field of the thermal bath, or solvent, changes appreciably with increasing concentration of the colloidal particles.

Identification of the partitioning of microstates in $C^*(q,\tau^*)$ has noteworthy consequences; One follows from extrapolation of power-laws fitted to A and $\tau_x$ (Fig. 1b) which find that $A \to \infty$, and $\tau_x \to 0$ at packing fractions $\phi_a$ and $\phi_b$ that equate, within experimental error, with the known freezing value, $\phi_f=0.494$ (Table 2). Instantaneous sampling ($\tau_x \to 0$) of the microstates is consistent with there being just one ($N \to 1$) accessible state where essentially all the acoustic energy ($A \to \infty$) resides. Here we arrive at a limit, attained by extrapolation beyond the actual data, that appears as unphysical ($A \to \infty$) as it is improbable ($S \to 0$), and may be seen merely as a fortuitous product of the analysis. Alternatively, one may consider this limit indicative of an entropic termination, $S \sim \ln(N) \to 0$, "at" the thermodynamic freezing point and, accordingly, identify the limit of thermodynamic stability of the suspension's fluid phase with the partitioning limit ($N \to 1$) of accessible microstates consistent with that phase. As such, the partitioning limit of the microstates, of the longitudinal momentum currents in this case, presents a definition of the freezing point.

To be clear, this determination of the freezing point of the suspension's fluid phase is based solely on the properties of that phase and contains no information about the other phase into which it transitions. However, accepting the above definition, it follows that an increase in $\phi$ beyond $\phi_f$ into the metastable, two-phase region necessitates collective, structural dynamics – as in caging – that now impact on the transverse momentum currents. Their ultimate partitioning effects separation of the crystal phase. Before this happens – while the suspension appears amorphous – the occurrence of collective dynamics in the metastable fluid means not all particle currents, $\mathbf{j}(q,t)$, are Gaussian. This, as mentioned under *Results and Analysis* and shown in detail in previous work [23], is seen by deviation of the CAF from SE decay for those spatial frequencies, q* around $q_m$, that manifest that collective dynamics.

So, traversal of the freezing point is also evident from a dynamical cross-over – a qualitative change in the decay of the CAF. How finely the cross-over brackets the above partitioning limit and thereby locates the freezing point is, of course, limited by the accuracy and resolution of the observations.

The more usual approach presumes all momenta have relaxed to the equilibrium, Maxwell-Boltzmann, distribution. From this position the thermodynamic limit of the fluid phase can be determined only by tying it to the crystal phase at the same pressure and free energy, or the entropy, in this case [3].

Finally, we suggest that the reduction in the number of microstates consistent with the one-phase thermodynamic macrostate is a general feature of a fluid's dynamics when approaching its freezing point. This may, however, not be apparent because the time scales of energy exchanges and structural relaxation, so advantageously separated in a suspension, overlap in atomic fluids. Nonetheless, recent MD simulations [28,29] of atoms with hard sphere and Lennard-Jones interactions close to their respective freezing points found long-time negative tails in the CAFs that could be approximated by SEs. So, for these conditions and more specifically for spatial frequencies, $q \approx q_m$, where, by virtue of de Gennes slowing [30], structural relaxation is slowest, sufficient delay occurs between energy exchanges and structural relaxation to expose the random sampling of the fluid's microstates. In all cases the decrease in decay times and stretching of the CAFs, respectively indicative of the decrease in the number and increase in spread of the amplitudes of those microstates, is clearly evident on the approach to the respective freezing points. However, the extant data is not sufficient for extrapolation and location of the partitioning limit with confidence. At the same time cross-overs from one-phase to two-phase regions, identified by emergence of caging mentioned above, were found to be consistent with the respective known freezing points.

*Summary/Conclusions*. – Scaling the delay time by that that characterises Brownian concentration fluctuations of a hard sphere suspension renders its current auto-correlation function independent of spatial frequency and expressible by a superposition of single exponential decays. From analyses of this quantity we conclude:

First, that the average decay time of the exponential decays reflects the number of microstates – the orthogonal longitudinal acoustic modes in the interstitial solvent, and thereby the entropy.

Second, that stretching of the CAF expresses the tendency of the frequency distribution of those acoustic modes to have a negative skewness.

Third, the partitioning limit of the microstates of the suspension's single phase fluid presents a definition of the freezing point.

Fourth, any increase in $\phi$ beyond this point into the two-phase region will necessarily lead to structures that must impact the transverse currents.

These findings ratify the statistical equivalence of the measured time average, of the CAF in this case, and that obtained from the average over the ensemble of microstates.

ACKNOWLEDGEMENTS. Peter Bourke, Gary Bryant, Peter Daivis, Martin Oettel, Miriam Klopotek. This work was financially supported by the DFG (Grant No. SCHO 1054/7-1)